\documentclass[aps,prd,twocolumn,preprintnumbers,nofootinbib,superscriptaddress,amsmath]{revtex4-2}
\usepackage[english]{babel}
\usepackage{bm}
\usepackage[utf8]{inputenc}
\usepackage[colorinlistoftodos, color=green!40, prependcaption]{todonotes}
\usepackage{amssymb}
\usepackage[pdftex, pdftitle={Article}, pdfauthor={Author}]{hyperref} 
\usepackage{placeins}

\begin{document}

\title{Evidence for dynamical dark energy from DESI-DR2 and SN data? A symbolic regression analysis}
\author{Agripino Sousa-Neto}
\email{agripinoneto@on.br}
\affiliation{Observatório Nacional, Rio de Janeiro - RJ, 20921-400, Brasil}
\author{Carlos Bengaly}
\email{carlosbengaly@on.br}
\affiliation{Observatório Nacional, Rio de Janeiro - RJ, 20921-400, Brasil}
\author{Javier E. Gonzalez}
\email{javiergonzalezs@academico.ufs.br}
\affiliation{Universidade Federal de Sergipe, São Cristóvão - SE, 49107-230, Brasil}
\author{Jailson Alcaniz}
\email{alcaniz@on.br}
\affiliation{Observatório Nacional, Rio de Janeiro - RJ, 20921-400, Brasil}
\date{\today}

\begin{abstract}
Recent measurements of Baryon Acoustic Oscillations (BAO) from the Dark Energy Spectroscopic Survey (DESI DR2), combined with data from the cosmic microwave background (CMB) and Type Ia supernovae (SNe), challenge the $\Lambda$-Cold Dark Matter ($\Lambda$CDM) paradigm. They indicate a potential evolution in the dark energy equation of state (EoS), $w(z)$, as suggested by analyses that employ parametric models. In this paper, we use a model-independent approach known as high performance symbolic regression (PySR) to reconstruct $w(z)$ directly from observational data, allowing us to bypass prior assumptions about the underlying cosmological model. Our findings confirm that the DESI DR2 data alone agree with the $\Lambda$CDM model ($w(z) = -1$) at the redshift range considered. Additionally, when combining DESI data with existing compilations of SN distance measurements, such as Patheon+ and DESY5, we observe no deviation from the $\Lambda$CDM model within $3\sigma$ (C.L.) for the interval of values of present-day matter density parameter $\Omega_m$ and the sound horizon at the drag epoch $r_d$ currently constrained by observational data. Therefore, similarly to the DESI DR1 case, these results suggest that it is still premature to claim statistically significant evidence for a dynamical EoS or deviations from the $\Lambda$CDM model based on the current DESI data in combination with supernova measurements.

\end{abstract}

\maketitle


\section{Introduction}
In recent decades, various cosmological observations have revealed that the Universe is undergoing acceleration \cite{Riess_1998,Perlmutter_1999,WMAP:2012nax,2020}. Within the framework of general relativity, one potential explanation for this phenomenon is the presence of a dark energy component. This component is characterized by a negative equation of state (EoS) parameter $w(z)$ and is interpreted as a Cosmological Constant $\Lambda$ ($w = -1$) in the standard cosmological model. In order to better understand the nature of such a component, significant efforts have been made to measure $w(z)$ and thus characterize its main physical properties. In this context, initiatives such as the Dark Energy Task Force (DETF)~\cite{DETF} outlined a roadmap to improve our understanding, classifying surveys into distinct stages. Among these are Stage IV surveys~(see, e.g., \cite{lsst,Euclid,SKA}), which include the Dark Energy Spectroscopic Instrument (DESI) \cite{desicollaboration2016desiexperimentisciencetargeting}.

The DESI collaboration, in its first data release of baryon acoustic oscillation (BAO) measurements, provided important results which challenge the $\Lambda$-Cold Dark Matter ($\Lambda$CDM) paradigm~\cite{desicollaboration2024desi2024vicosmological}, as further intensified by subsequent works~\cite{ChanGyungPark2024,yin2024cosmiccluesdesidark,bousis2024hubbletensiontomographybao,luongo2024modelindependentcosmographicconstraints,cortes2024interpretingdesisevidenceevolving}. Assuming that the dark energy EoS is given by the CPL parameterization~\cite{2001IJMPD..10..213C,2003PhRvL..90i1301L}, $w(z) = w_0 + w_az/(1+z)$, some analyses of the DESI-BAO data, when combined with various Type Ia Supernova (SNe) datasets, such as the Dark Energy Survey Supernova 5-Year compilation (DESY5) \cite{descollaboration2024darkenergysurveycosmology}, Union3~\cite{rubin2024unionunitycosmology2000}, and Pantheon+~\cite{Scolnic_2022}, in addition to Cosmic Microwave Background (CMB) priors from Planck \cite{2020}, showed a preference for {a dynamical dark energy (with $w_0 > -1$ and  $w_a < 0$)} compared to $\Lambda$CDM (see also \cite{Efstathiou:2024xcq,Roy_Choudhury_2024} for different perspective on this discussion). Additionally, this conclusion is supported by analyses that explore different EoS parameterizations~\cite{Giare:2024gpk,wolf2024,rebouças2024,bhattacharya2024,ramadan2024,Ghosh:2024kyd}, with the best-fit model provided by the BA parameterization~\cite{Giare:2024gpk}, $w(z)=w_0 + w_az(1+z)/(1+z^2)$~\cite{Barboza:2008rh}.

Another way to approach the cosmological data is through non-parametric analysis \cite{Purba2024,jiang2024,Ghosh2024,dinda2024modelagnosticassessmentdarkenergy}, {which can determine whether there is any evolution of the dark energy EoS with redshift without assuming a specific functional form for $w(z)$. Methods} like  Gaussian Processes (GP)\cite{RasmussenW06,Seikel_2012} and Genetic Algorithms (GA)\cite{Bogdanos_2009,Arjona_2020} have been widely used for this purpose, providing a flexible framework for reconstructing cosmological functions without relying on specific cosmological models. {Some non-parametric reconstructions of the DESI-BAO measurements, along with the aforementioned SNe data and CMB priors~\cite{dinda2024modelagnosticassessmentdarkenergy}, exhibited a smaller tension with the $\Lambda$CDM model  than those reported in \cite{desicollaboration2024desi2024vicosmological} assuming the CPL model.}

The goal in this work is to employ a novel model-independent approach, namely the Symbolic Regression using the High-Performance Symbolic Regression in Python and Julia library\footnote{https://github.com/MilesCranmer/PySR} (PySR) \cite{cranmer2023interpretable}, on the DESI-BAO and SN data in order to reconstruct the dark energy EoS. Hence, we can check whether there is any statistically significant departure of the $w=-1$ prediction from the standard model, which would help supporting or challenging previous results in the literature.

In this paper, we first review the relevant cosmological equations to our study in Section \ref{CF}, including the key concepts under consideration. Section \ref{Data} describes the data used in our analysis, detailing the specifics of the DESI, Patheon+, and DESY5 datasets. In Section \ref{SR}, we discuss the theoretical framework of symbolic regression, focusing on its application to cosmological data. Section \ref{Results} and Appendices A, B and C present the results of our analysis. Finally, Section \ref{Conclusion} summarizes our main conclusions.

\section{Basic Equations}\label{CF}

We consider a spatially flat universe and evolving dark energy within the general-relativistic framework of a statistically homogeneous and isotropic Universe, according to the Friedmann-Lemaître-Robertson-Walker metric. Under these assumptions, the cosmic expansion rate, described by the Hubble parameter, $H(z)$, is given by
\begin{equation}
\frac{H(z)^2}{H_0^2}= {\Omega_{m}(1+z)^3} + \Omega_{DE} \exp\left[3\int_{0}^{z} \frac{1+w(z')}{1+z'}dz' \right],
 \label{eq:wideeq}
\end{equation}
where $H_0$ is the Hubble constant, quantifying the current expansion
rate of the Universe, $\Omega_{m}$ is the present-day total matter density parameter, which includes both dark and baryonic matter, and $\Omega_{DE}$ is the current dark energy density parameter. With the assumption of a spatially flat universe, the density parameters satisfy the relation $\Omega_{m}+\Omega_{DE}=1$. 

One can invert the above equation to derive an expression for the dark energy EoS, $w(z)$, in terms of the $H(z)$, and its first derivative,
\begin{equation}\label{eq:wzinHz}
    w(z)=-\frac{1}{3}\frac{2(1+z)HH'-3H^2}{H_0^2(1+z)^3\Omega_m - H^2}\;,
\end{equation}
which can be rewritten as~\cite{dinda2024modelagnosticassessmentdarkenergy}
\begin{equation}\label{dinda&Maartins}
w(z) = -1 -  \frac{2(1+z)\tilde{D}'_H(z) + 3\gamma(1+z)^3\,\tilde{D}^{3}_H(z)}{3\big[\tilde{D}_H(z) - \gamma(1+z)^3\,\tilde{D}^{3}_H(z)\big]}\;.
\end{equation}
In the above expression, $\tilde{D}_H(z)=c/H(z)r_d$
and
\begin{equation}
    \gamma = \frac{\Omega_{\rm m} H_0^2 r_d^2}{c^2} \;,
\label{eq:defn_gamma}
\end{equation}
where $H_0=100h\,\text{km s}^{-1} \, \text{Mpc}^{-1}$, $c$ is the speed of light and {$r_d$ is the sound horizon at the drag epoch}. Unless stated otherwise, we hereafter assume the latter two quantities to be $r_d=(101 \pm 2.3) \, h^{-1}~\text{Mpc}$, as reported by \cite{Verde_2017}, and $\Omega_m=0.3$. Note that $D_M$ is related to $D_H$ by $D'_M=c/H(z)=D_H$, allowing us to rewrite Eq.~\eqref{dinda&Maartins} as  
\begin{equation}\label{dinda&Maartins2}
w(z) = -1 -  \frac{2(1+z)\tilde{D}''_M(z) + 3\gamma(1+z)^3\,\tilde{D}'^{3}_M(z)}{3\big[\tilde{D}'_M(z) - \gamma(1+z)^3\,\tilde{D}'^{3}_M(z)\big]}\;.
\end{equation}
For analyses involving SNe data, $\tilde{D}_M$ is calculated using the second equation above, which relates it to the luminosity distance, $D_L$, via the absolute magnitude, $M$, and the apparent magnitude $m$: 
\begin{equation}
\label{DLDM}
\tilde{D}_M = \frac{D_L}{(1+z)r_d} = \frac{10^{\frac{m - M - 25}{5}}}{(1 + z) r_d}\;.
\end{equation}

To obtain the EoS using Eq.~\eqref{dinda&Maartins}, it is necessary to reconstruct $\tilde{D}_H$ along with its first derivative. In contrast, when computing the EoS in terms of $\tilde{D}_M$ using Eq.~\eqref{dinda&Maartins2}, both the first and second derivatives are required.

\section{Data}\label{Data}
This section describes the observational datasets used in this work, including BAO measurements from DESI DR2 \cite{desicollaboration2025desidr2resultsii} and SNe data from the Pantheon+ \cite{Scolnic_2022} and DESY5~\cite{descollaboration2024darkenergysurveycosmology} samples.
\subsection{Baryonic acoustic oscillations}
The DESI-BAO measurements are expressed by three ratios: the three-dimensional BAO mode ($D_V / r_d$), the transverse mode ($D_M / r_d$), and the radial mode ($D_H / r_d$). In this work, we focus on the transverse and radial modes. An important point is that to reconstruct $w(z)$, we use only the measurements $D_H / r_d$ or $D_M / r_d $ in our analyses, depending on the equation used (Eq.\eqref{dinda&Maartins} or Eq.\eqref{dinda&Maartins2}). For this reason, we do not use $D_V / r_d$ from DESI data. For analyses using only DESI data, we use exclusively $D_H / r_d $ because the large uncertainties involved in the second derivative in Eq.\eqref{dinda&Maartins2} make it non plausible to reconstruct $ w(z)$ using only DESI with $ D_M / r_d$ measurements.

The radial comoving distance can be related to the transverse BAO mode measurement by
\begin{equation}
    \frac{D_M}{r_d}= \frac{D_A(1+z)}{r_d},
\end{equation}
where $D_A$ is the angular diameter distance. Similarly, the radial BAO mode can be written as
\begin{equation}\label{eq:DH}
   \frac{D_H(z)}{r_d}  = \frac{c}{H(z)r_d}.
\end{equation}
The BAO measurements obtained by the DESI collaboration are summarized in Table \ref{tab:table1}. There, we present the measurement values for different
tracers, namely the Luminous Red Galaxy (LRG), divided into two samples (LRG1, LRG2), the combined LRG3 and Emission Line Galaxy (ELG1), ELG2, Lyman-$\alpha$ forest, and the QSO sample. The first column indicates the tracers, the second one displays the redshift range of the tracers, while the third column shows the effective redshift, $z_{eff}$. The fourth and the last columns list the $D_H(z)/r_d$ and $D_M(z)/r_d$ ratios, along with their 1$\sigma$ confidence limits.
\begin{table}[!ht]

\begin{ruledtabular}
\begin{tabular}{lcccc}
tracer &redshift& $z_{eff}$&  $D_H/r_d$& $D_M/r_d$
\\
\colrule
LRG1&$0.4-0.6$ & $0.510$ &  $21.863\pm0.425$ & $13.588\pm0.167$ \\
LRG2&$0.6-0.8$  & $0.706$ &  $19.455\pm0.330$ &$17.351\pm0.177$\\
LRG3+ELG1&$0.8-1.1$  & $0.934$ & $17.641\pm0.193$&$21.576\pm0.152$ \\
ELG2&$1.1-1.6$  & $1.321$ &  $14.176\pm0.221$ & $27.601\pm0.318$\\
QSO&$0.8-2.1$ & $1.484$ & $12.817\pm0.516$&$30.512\pm0.760$\\
Ly$\alpha$&$1.8-4.2$& $2.330$ & $8.632\pm0.101$&$38.988\pm0.531$\\
\end{tabular}
\caption{\label{tab:table1}%
BAO measurements from the DESI (DR2) collaboration \cite{desicollaboration2025desidr2resultsii}.}
\end{ruledtabular}
\end{table}
\subsection{Type Ia Supernova}
\subsubsection{Pantheon+}
The Pantheon+ dataset encompasses 20 supernova compilations spanning a redshift range from $0.00122$ to $2.26137$ \cite{Scolnic_2022}. This dataset includes 1701 light curves corresponding to 1550 spectroscopically confirmed SNe. At low redshifts, $z < 0.01$, 111 SNe have been removed in order to prevent a strong peculiar velocity dependence. The 1590 light curves in our final sample span a redshift range from $0.01016\leq z \leq 2.26137$.  The data and the covariance matrix used in this work are available in the GitHub repository\footnote{\url{https://github.com/PantheonPlusSH0ES/DataRelease/tree/main/Pantheon\%2B_Data}}.

\subsubsection{Dark Energy Survey Supernova 5-Year (DESY5) }

The DESY5 supernova sample comprises 1635 supernovae observed by the DES collaboration, spanning a redshift range from 0.0596 to 1.12, along with 194 high-quality external supernovae at redshifts below 0.1, for a total of 1829 supernovae. After excluding supernovae with magnitude errors $\delta m > 1$, which are unlikely to be classified as Type Ia based on the DES photometric data, our final sample consists of 1754 supernovae. The complete catalog is accessible through the DES collaboration's GitHub repository \footnote{ \url{https://github.com/des-science/DES-SN5YR}.}.

\section{Symbolic regression}\label{SR}
Symbolic Regression (SR) is a supervised learning task in which the model space consists of analytic expressions \cite{koza1994genetic,schmidt2009distilling}. It is commonly tackled through a multi-objective optimization framework, aiming to minimize prediction error and model complexity simultaneously. Instead of adjusting specific parameters within an overparameterized general model, this method explores the space of simple analytic expressions to identify accurate and interpretable models. SR can also be employed as a non-parametric method, particularly in reconstructing cosmological observables.

Among the various SR libraries available in the literature, such as Eureqa \cite{schmidt2009distilling}, GPLearn\footnote{\url{https://github.com/trevorstephens/gplearn}}, and AI Feynman \cite{udrescu2020ai}, we will utilize PySR \cite{cranmer2023interpretable}. PySR is a powerful open-source library designed to efficiently discover symbolic models using a combination of evolutionary algorithms and gradient-based optimization. 
\subsection{Operators Used in PySR}
In symbolic regression with PySR, various operators construct and evolve symbolic expressions. We use a predefined set of unary and binary operators to explore possible solutions and identify the best-fitting models, as shown in the table below.
\begin{table}[!ht]
\label{tab:SR}
\begin{ruledtabular}
\begin{tabular}{lc}
unary & binary
\\
\colrule
log & $-$  \\
sqrt  & $+$  \\
$\exp$  & $*$  \\
cube   & $/$  \\
square  & pow \\

\end{tabular}
\caption{ }

\end{ruledtabular}
\end{table}

These operators are used to construct the symbolic expressions that are evaluated and optimized to find the most accurate and interpretable models for the given data.

\subsection{Loss Function}
Loss functions quantify how well a model can simulate the intended result. In order to account for the uncertainties in the data and make sure that the model assigns the right weight to each observation, we use a custom loss function\footnote{See more details and another PYSR loss functions in \url{https://ai.damtp.cam.ac.uk/symbolicregression/stable/losses/}}. We adopt the loss function defined as
\begin{equation}
\mathcal{L} = \frac{1}{n} \mathbf{r}^T \mathbf{C}^{-1} \mathbf{r}\;,
\end{equation}
where $\mathbf{r} = x - y$ is the residual vector, with $x$ representing the predicted value and $y$ denoting the true value; $\mathbf{C}^{-1}$ is the inverse covariance matrix; and $n$ is the number of points in the sample, used to normalize the loss. This approach allows the model to better account for data quality and focus on more reliable observations during optimization.
\subsection{Model Selection}\label{sec:model_selection}
In this work, we use the ``best'' selection criterion from PySR, which selects the equation with the best trade-off between error and complexity. First, PySR defines a threshold for the error, calculated as 1.5 times the minimum error among all models. Only equations with an error below this threshold are considered. Then, the equation with the highest score is selected among the filtered models. The resulting expression represents the best empirical description of the data within the considered search space\footnote{Other available selection criteria can be found in the official PySR documentation: \url{https://ai.damtp.cam.ac.uk/pysr/api/}. The final selected expressions after the process are listed in Appendix~\ref{appendix:hf}.}.

Although these expressions exhibit good statistical performance, they do not inherently guarantee physical meaning, as they are guided solely by data fitting and structural simplicity. For this reason, it is sometimes necessary to impose physical constraints, such as $D_M = 0$ at $z = 0$, to guide the algorithm toward learning representations consistent with physical expectations.

\section{Results}\label{Results}
In what follows, we present the reconstructed results obtained using the method described in the previous section. First, we show the results for the DESI DR2 dataset. Subsequently, we combine the DESI-BAO data with the two supernova datasets described earlier.
\begin{figure*}[hbt!]
    \centering
     \includegraphics[scale=0.26]{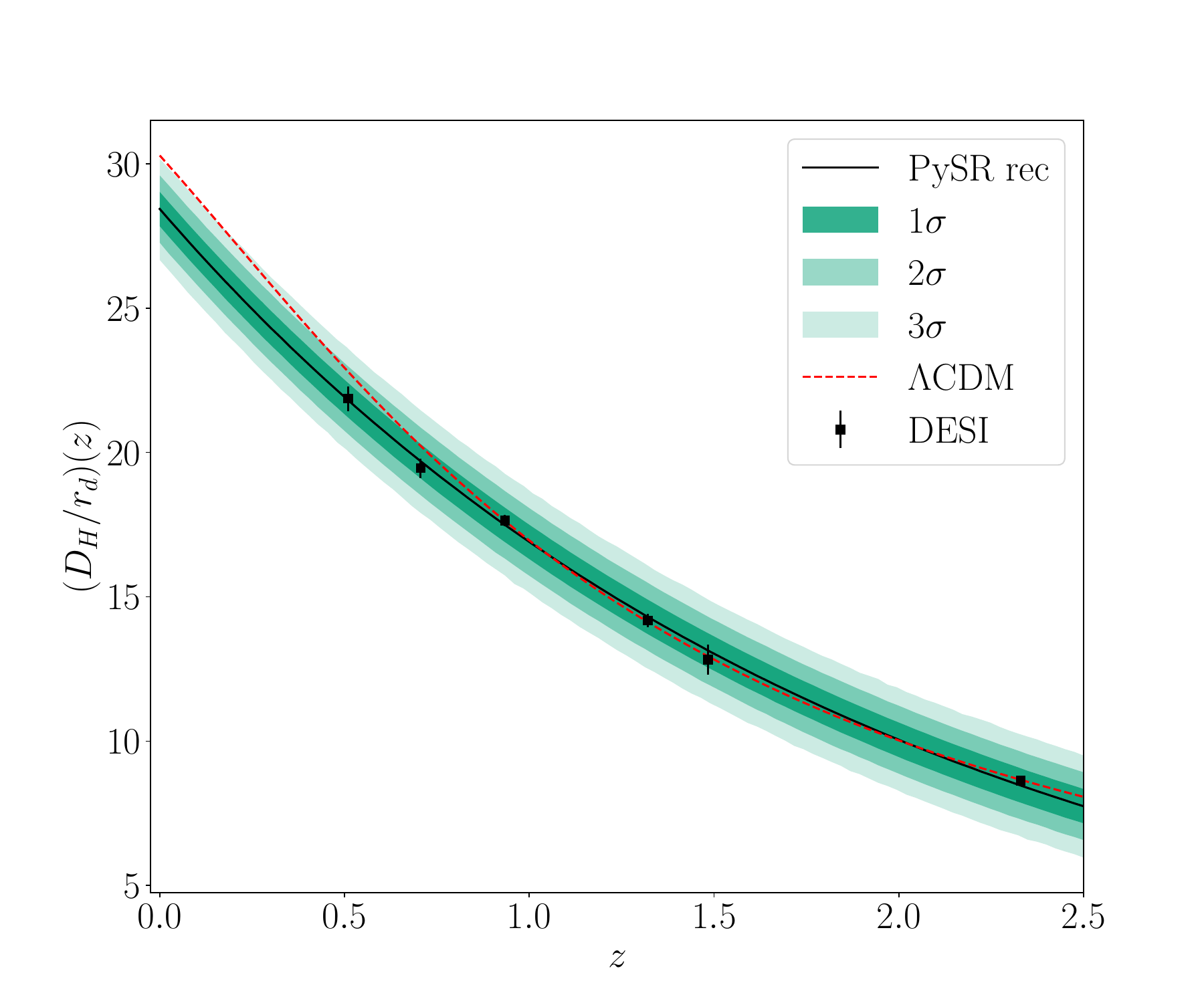}
    \includegraphics[scale=0.26]{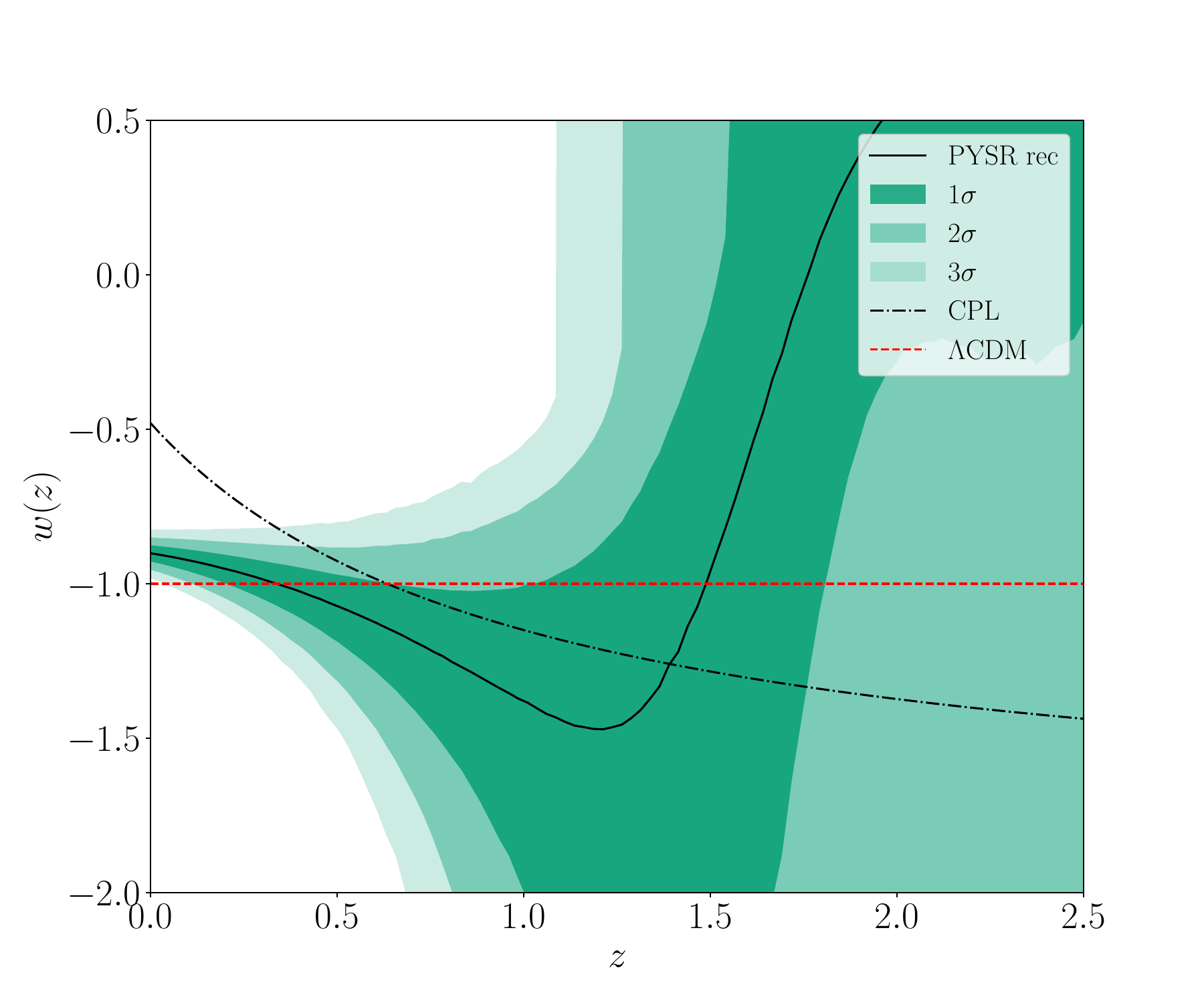 }
    \caption{The reconstruction results for $D_H/r_d$ DESI-only. The black line represents the mean for the PySR reconstruction; the green bands indicate the CLs. The red dashed line corresponds to the $\Lambda$CDM theoretical curve, while the black dashed line represents the CPL model with the $w_0-w_a$ DESI-BAO results. \textbf{Left:} $D_H/r_d$ reconstruction with PySR. The black points correspond to the DESI-BAO dataset, with respective uncertainties.
    \textbf{Right:} The $w(z)$ derived using Eq.~\eqref{dinda&Maartins} and the reconstructed $D_H/r_d$, assuming $r_d=101\pm2.3h^{-1}$ Mpc, as reported by \cite{Verde_2017}, and $\Omega_m=0.3$.} 
    \label{fig:DHrdfull}
\end{figure*}
\subsection{DESI-only}

For completness, we begin our analysis by reconstructing the quantity $D_H/r_d$ from the DESI-BAO dataset. The best-fit equation obtained by PySR to describe these data is given by
\begin{equation}\label{eq:DHbest} 
\frac{D_H}{r_d}= 28.4 \exp(-0.519z)\;, 
\end{equation} 
which corresponds to the model selected using the ``best” criterion described in Section~\ref{sec:model_selection}. Other candidate equations generated during the training, along with their complexity and loss, are listed in Table~\ref{table:DESI-only} in Appendix~\ref{appendix:hf}.

The results is displayed in the left panel of Figure~\ref{fig:DHrdfull}, along with the $1\sigma$, $2\sigma$, and $3\sigma$ confidence levels (CL). These CL are estimated using the Monte Carlo method, assuming a Gaussian distribution centered on the best-fit equation provided by PySR, with standard deviations given by the reconstruction uncertainties.

As for the dark energy EoS, $w(z)$, we can compute it by plugging the reconstructed $D_H(z)/r_d$ given by Eq.~\eqref{eq:DHbest} and its derivative into  Eq.~\eqref{dinda&Maartins}, as presented in the right panel of Figure \ref{fig:DHrdfull}. The dashed red line represents the $\Lambda$CDM scenario, $w(z)=-1$, while the black curve corresponds to the CPL parametrization with the values $w_0 = -0.48$ and $w_a = -1.34$ from the DESI-BAO results \cite{desicollaboration2025desidr2resultsii}. The green-shaded regions correspond to the reconstruction's 1$\sigma$ to 3$\sigma$ CL (from darker to lighter shade, respectively), also obtained using the Monte Carlo method. At $1\sigma$ level, this analysis shows good agreement with the $\Lambda$CDM prediction up to $z \approx 0.66$. 

In our final analysis using DESI-only data, we examine the sensitivity of $w(z)$ to variations in the parameters $\Omega_m$ and  $r_d$, which enters in the calculation of $\gamma$ -- Eq. (\ref{eq:defn_gamma}). Specifically, we explore $\Omega_m$ values of 0.25, 0.30, and 0.35, while testing $r_d$ at $90.9 \,h^{-1} \, \text{Mpc}$, $101 \,h^{-1} \, \text{Mpc}$, and $ 111.1 \,h^{-1}\,\text{Mpc}$, based on values reported by \cite{Verde_2017} (see also \cite{Lemos:2023qoy} for model-independent estimates of $r_d$ from angular BAO measurements). The results, detailed in Appendix \ref{append:gamma}, demonstrate how changes in these parameters influence the reconstruction of $w(z)$, highlighting the sensitivity of the equation of state to the cosmological parameters in Eq. (\ref{eq:defn_gamma}). In particular, we find deviation from the $\Lambda$CDM model at $3\sigma$ for values of $\Omega_m \leq 0.25$ and $r_d \leq 90.9 \,h^{-1} \, \text{Mpc}$.

\subsection{DESI + Pantheon+}\label{desipant}

Our first analysis combining the DESI dataset with SNe data focuses on the Pantheon+ dataset. We use the apparent magnitude, $m$, from the Pantheon+ compilation to reconstruct the quantity $D_M / r_d$ fixing $M = -19.253$~\cite{Riess:2021}. Then, we compute the normalized comoving distance ($D_M$) of the SNe assuming  Eq.~\eqref{DLDM} and $r_d = 138.28\pm 3.1 \, \mathrm{Mpc}$, which corresponds to the $r_d$ value given by \cite{Verde_2017} combined with the $H_0$ value reported in \cite{Riess:2021} -- which is consistent with the $M$ value previously assumed. Finally, we  compute the corresponding $D_M/r_d$ values from the SNe apparent magnitude measurements, so that they can be combined with the $D_M/r_d$ from the DESI-BAO measurements. This scenario is denoted as DESI + Pantheon+.
 
It is important to note that the SNe data do not constrain $H_0$ and $M$ independently, but the combination ${\mathcal M=}M-5\log H_0$. In this sense, the values of $M$ and $H_0$ must satisfy this relation to be compatible for a specific SNe data set. Thus, Eq.~\eqref{DLDM} can be rewritten in terms of ${\mathcal{M}}$ and $H_0$ to obtain:

\begin{equation}
    \tilde{D}_M \propto \frac{10^{-{\mathcal{M}}/5}/H_0}{r_d}.
\end{equation}
Since an uncalibrated sound horizon scale is used (Ref. \cite{Verde_2017}), where $r_d \propto 1/H_0$,  our SNe data expressed in terms of $\tilde{D}_M$ and the parameter $\gamma$ in Eq.~\eqref{eq:defn_gamma} are independent of the $H_0$ value.  Consequently, the reconstruction of the dark energy EoS in Eq.~\eqref{dinda&Maartins2} does not depend on the value of $H_0$ and $M$.

The {best} equation derived by PySR for the $D_M/r_d$ reconstruction, in the first scenario, is given by 
\begin{equation}\label{bestPantheon}
  \frac{D_M}{r_d} = 33.3  z  \exp(-0.274z)\;.
\end{equation}

\begin{figure*}[!ht]
    \centering
    \includegraphics[scale=0.26]{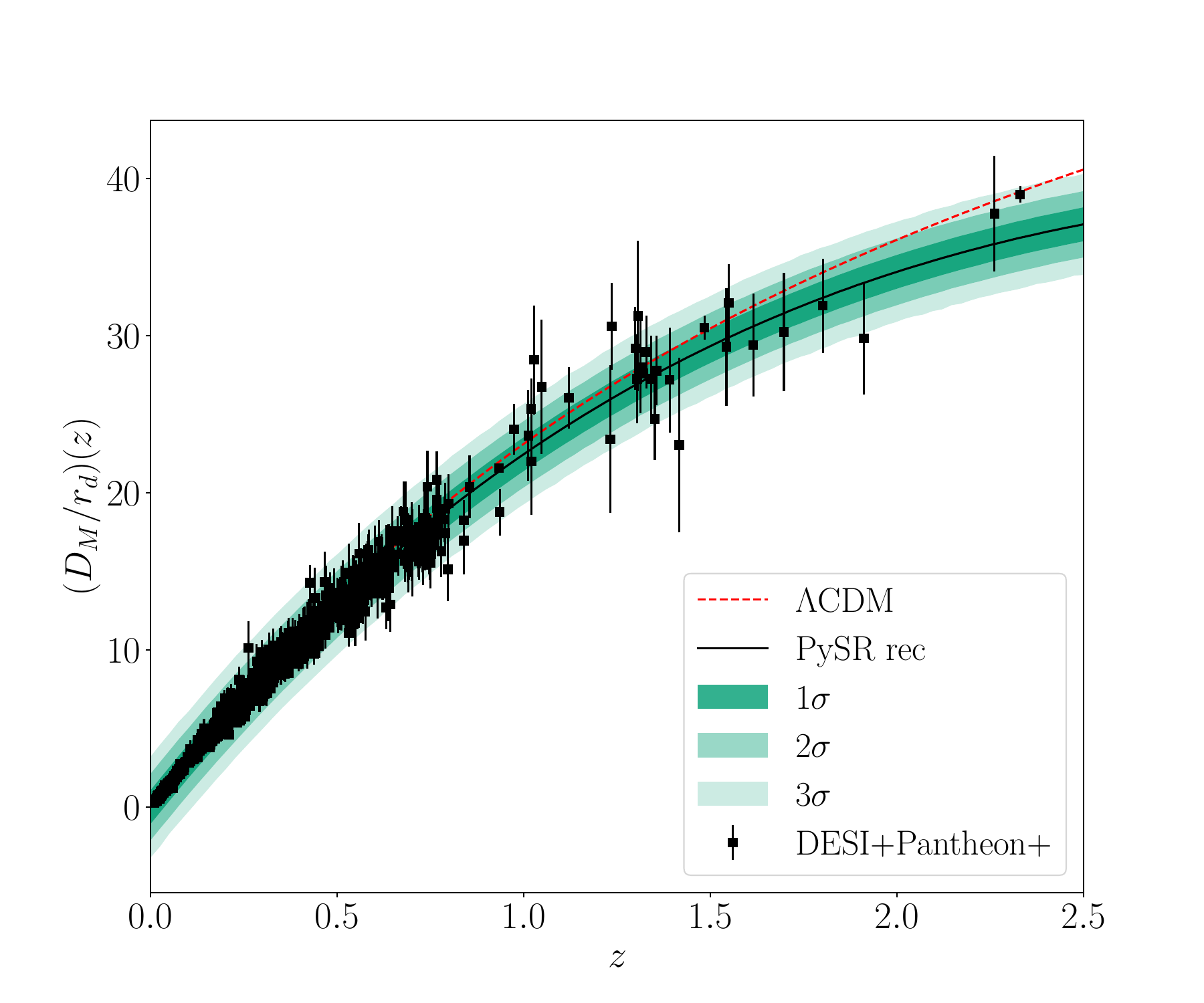}
    \includegraphics[scale=0.26]{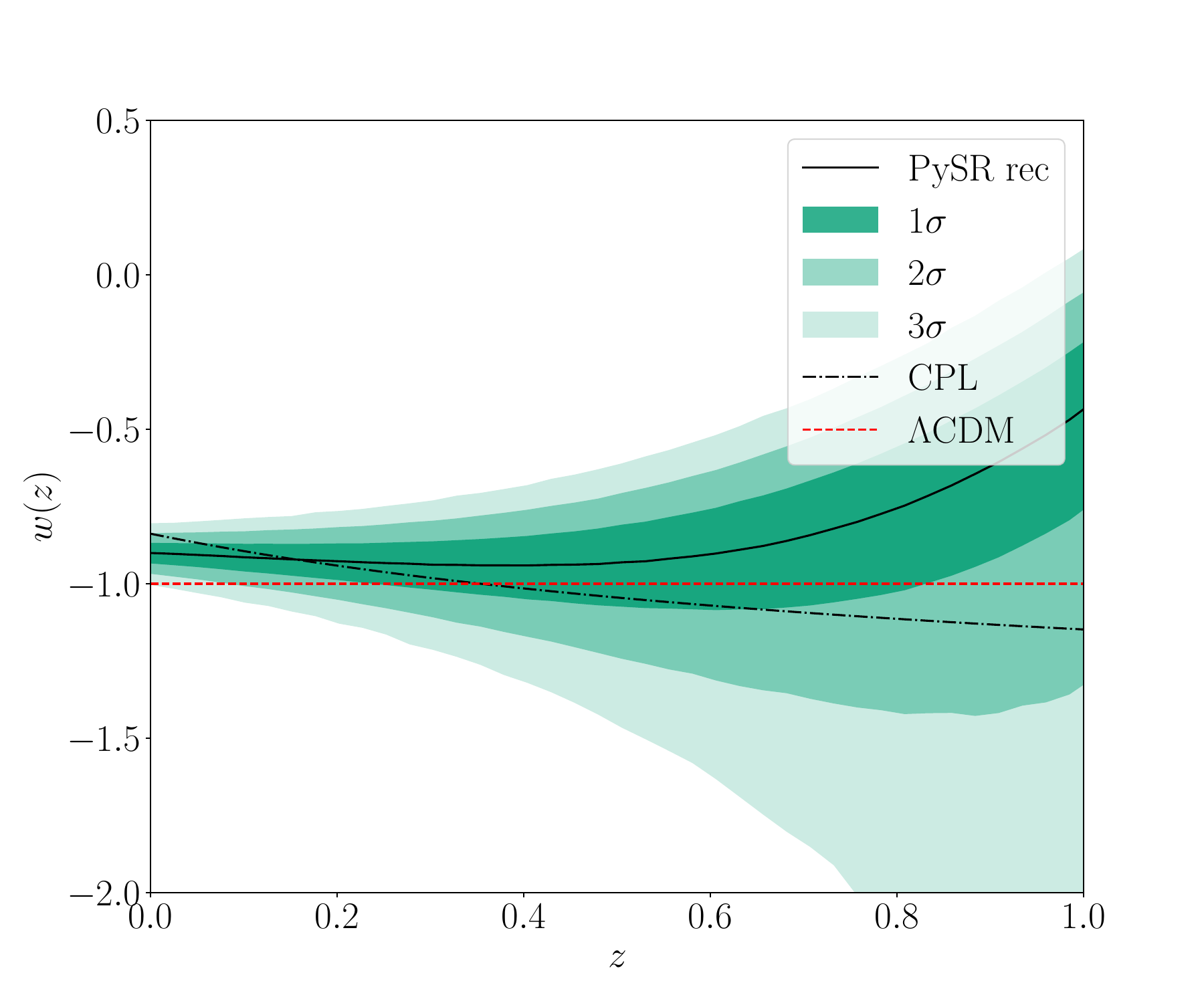}
\caption{The reconstruction results for $D_M/r_d$ using DESI + Pantheon+ dataset. The black line represents the mean for the PySR reconstruction; the green bands indicate the confidence levels. The red dashed line corresponds to the $\Lambda$CDM theoretical curve. \textbf{Left:} $D_M/r_d$ reconstruction with PySR. The black points correspond to the dataset, with respective uncertainties. 
\textbf{Right:} The $w(z)$ derived using Eq.~\eqref{dinda&Maartins2} and the reconstructed $D_M/r_d$, assuming $r_d=101\pm2.3h^{-1}$ Mpc, as reported by \cite{Verde_2017}, and $\Omega_m=0.3$. While the black dashed line represents the CPL model with the DESI + CMB + Pantheon+ results \cite{desicollaboration2025desidr2resultsii}.}
\label{fig:pantheonDR2}
\end{figure*}

The left panel of Figure \ref{fig:pantheonDR2} shows the reconstruction of ${D_M(z)}/{r_d}$ for the DESI + Pantheon+ dataset. The reconstructed $D_M/r_d$ values agree with the $\Lambda$CDM model within $1\sigma$ for $z \lesssim 1.7$, but deviate from it at larger $z$, reaching the $3\sigma$ region. A probable reason for this deviation is the inclusion of Pantheon+ data points at $z > 1$, where larger uncertainties and possible systematics are more prevalent. In contrast, DESI data alone show a better agreement with the $\Lambda$CDM model in high-redshift regions.

The result of the $w(z)$ reconstruction for the DESI + Pantheon+ is shown in the right panel of Figure ~\ref{fig:pantheonDR2}. The reconstructions of $w(z)$ for different values of $\gamma$ are shown in Appendix~\ref{append:gamma}. We find that both the $\Lambda$CDM model and the CPL parametrization with the DESI + CMB + Pantheon+ results \cite{desicollaboration2025desidr2resultsii} fall within the $3\sigma$ bounds of the reconstructions, considering their large uncertainties. 

\subsection{DESI + DESY5}

The procedure for reconstructing the ratio $D_M/r_d$ follows the same methodology as in the analysis presented in the section~\ref{desipant}. However, instead of using the Pantheon+ SNe dataset, we utilize the DESY5 dataset, {whose absolute magnitude is consistent with a Hubble constant calibration of $H_0 = 70 \, \mathrm{km \, s^{-1} \, Mpc^{-1}}$}. 

The resulting data for  DESI + DESY5, is used to derive a new relationship for $D_M / r_d$. The {best} equation obtained from PySR for this dataset is 
\begin{equation}\label{bestDESY5}  
\frac{D_M}{r_d} = 30.1  z \exp(-0.262z). 
\end{equation}  

\begin{figure*}[!ht]
    \centering
    \includegraphics[scale=0.26]{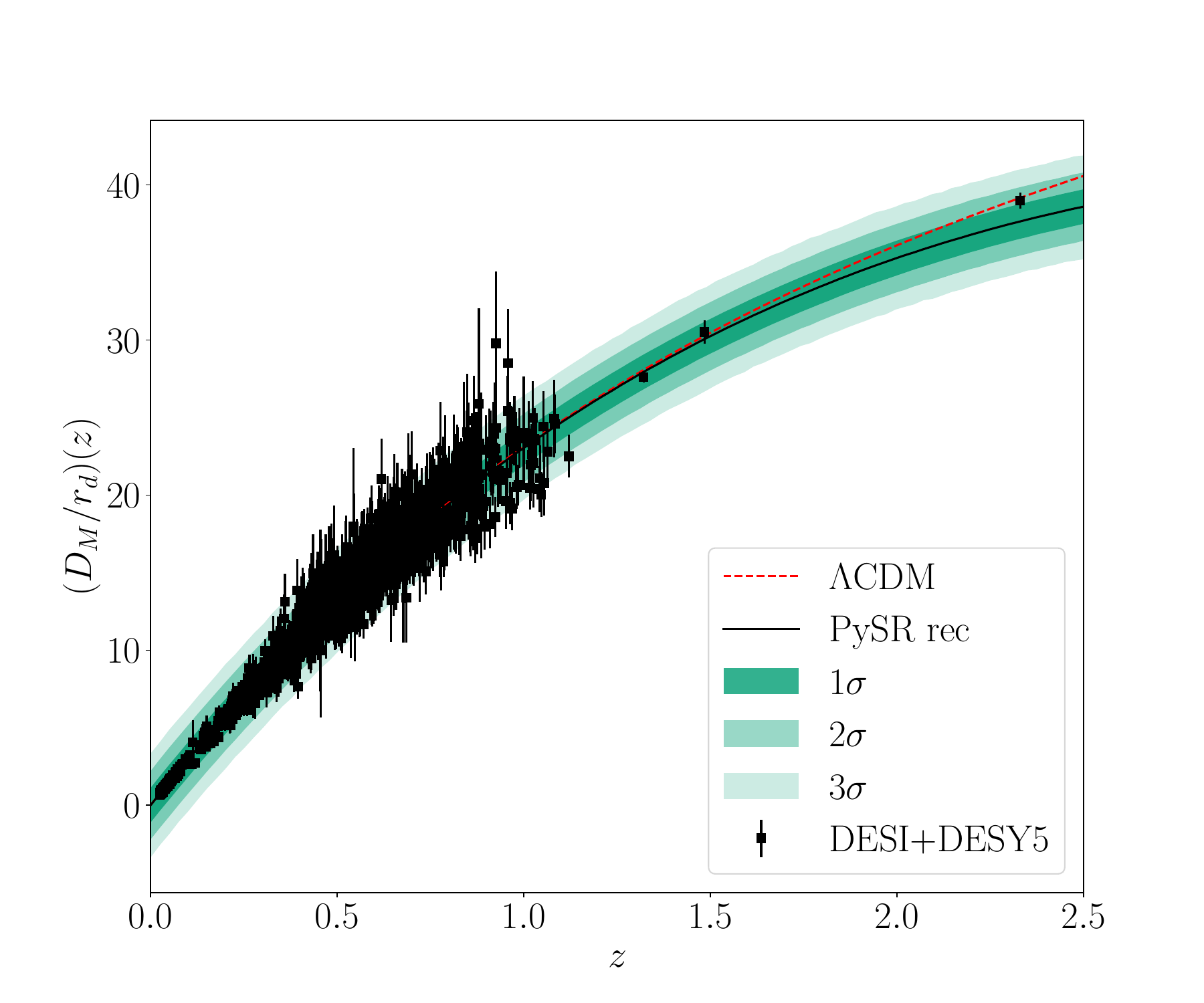}
    \includegraphics[scale=0.26]{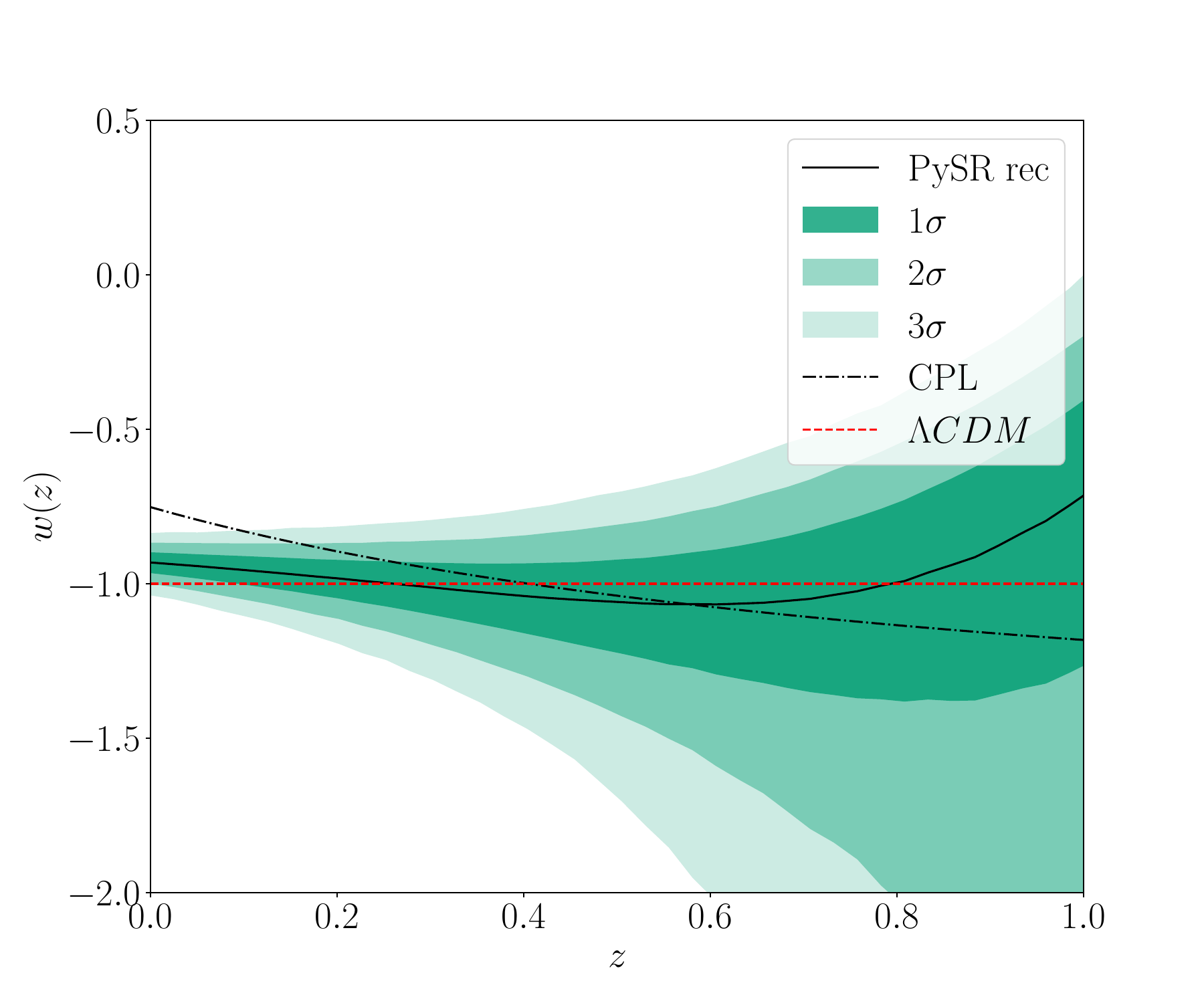}
\caption{ The same as Figure~\ref{fig:pantheonDR2}, but using the DESI + DESY5, and a CPL model with the DESI + CMB + DESY5 results \cite{desicollaboration2025desidr2resultsii} .}
    \label{fig:DESY5DR2}
\end{figure*}

The reconstruction of $D_M/r_d $ for DESI + DESY5 is shown in the left panel of Figure \ref{fig:DESY5DR2}. The reconstruction agrees with  $\Lambda$CDM within the 2$\sigma$ CL, as the dashed red line indicates. Additionally, the DESI + DESY5 reconstruction of $D_M/r_d$ aligns more closely with $\Lambda$CDM than the DESI + Pantheon+ case.

The resulting $w(z)$ {for the scenario DESI + DESY5} is displayed in the right panel of Figure \ref{fig:DESY5DR2}, which is in agreement with $\Lambda$CDM within $2\sigma$, as indicated by the dashed red line. Nevertheless,  the the best-fit values of the CPL parametrization for DESI + CMB + DES-SN5YR (see e.g. \cite{desicollaboration2025desidr2resultsii}) deviates from the reconstruction from $z=0$ up to $z \sim 0.2$, falling outside the $3\sigma$ region.
\section{Conclusion}\label{Conclusion}

Given its potential consequences for cosmology, DESI's result showing a preference for a time-dependent dark
energy EoS has motivated several analyses using parametric and non-parametric methods. In this paper, we investigated this issue from a High-Performance Symbolic Regression (PySR) reconstruction of $w(z)$. The reconstruction is first performed using the DESI $D_H/r_d$ measurements and then refined by combining DESI $D_M/r_d$ with SNe data. Our primary result shows that asserting that they favor any particular cosmological model is premature. Such a conclusion is due to the large uncertainties derived from these data, which limit the precision needed to support a specific model definitively.

The EoS, $w(z)$, shows significant sensitivity to the cosmological parameters entering the $\gamma$ expression -- Eq. (\ref{eq:defn_gamma}). Variations in both $\Omega_m$ and $r_d$ lead to noticeable changes in the reconstruction of $w(z)$, highlighting the dependence of the equation of state on these key cosmological parameters. However, within the explored range of $\Omega_m$ and $r_d$, the reconstructed EoS remains consistent with $\Lambda$CDM, except for the lower values of $\Omega_m$ and $r_d$, which show mild deviations.

The $w(z)$ for this scenario shows no statistically significant departure of $\Lambda$CDM within $3\sigma$ confidence level; however, the large error bars prevent the exclusion of alternative models

Concerning the DESI + DESY5, the reconstructed $D_M/r_d$ values also agree with the $\Lambda$CDM model within $1\sigma$. Since the DESY5 dataset only extends up to $z=1.12$, the following points in the reconstruction come from DESI-BAO data, which agree well with $\Lambda$CDM. This alignment likely caused the reconstruction to fit more closely to $\Lambda$CDM than the previous scenario with Pantheon+ data. The $w(z)$ reconstruction from the DESY5 dataset agrees with $\Lambda$CDM within $2\sigma$.

Finally, it is worth mentioning that our results aligned with the overall conclusion obtained through other non-parametric analyses (see, e.g., \cite{dinda2024modelagnosticassessmentdarkenergy}), although the evolution of $w$ with redshift differs from other studies. {In particular, the statistical level of compatibility between our reconstruction and the $\Lambda$CDM model is $3\sigma$, which matches the statistical significance shown in the DESI DR2 + CMB + Union3 analysis using Gaussian Process regression (Figs. 9 and 10)} \cite{desicollaboration2025desidr2resultsii}.

\begin{acknowledgments}
ASN is supported by the Coordena\c{c}\~ao de Aperfei\c{c}oamento de Pessoal de N\'ivel Superior (CAPES). CB acknowledged financial support from Funda\c{c}\~ao de Amparo \`a Pesquisa do Estado do Rio de Janeiro (FAPERJ). JSA is supported by CNPq grant No. 307683/2022-2 and FAPERJ grant No. 259610 (2021). This work was developed thanks to the use of the National Observatory Data Center (CPDON).
\end{acknowledgments}

\appendix
\section{Top-performing equations}\label{appendix:hf}
This appendix presents the most notable symbolic expressions generated by PySR during our analysis. Each entry showcases a unique solution that achieved high accuracy in reconstructing the cosmological parameters under different datasets and scenarios. 

The results are summarized in the following tables:
\begin{itemize}
    \item Table \ref{table:DESI-only}: DESI-only;
    \item Table \ref{table:DESI+PantheonPlus}: DESI combined with Pantheon+;
    \item Table \ref{table:DESI+DESY5}: DESI combined with DESY5.
\end{itemize}
Each table includes the complexity of the expression, the corresponding loss function, and the mathematical equation representing the solution. The best equations selected are in bold. These results highlight the diversity of the expressions found by PySR. 

\begin{table}[!ht]
    \centering
    \caption{DESI-only}
    \resizebox{\columnwidth}{!}{%
    \begin{tabular}{|c|c|l|}
        \hline
        \textbf{Complexity} & \textbf{Loss} & \textbf{Equation} \\ \hline
        $1$ & $2519.06$ & $y =   2.8069465$ \\
        $2$ & $120.73$ & $y = \sqrt{276.74304}$ \\
        $3$ & $91.5$ & $y = 17.705526 - z$ \\
        $4$ & $1.22$ & $y = (5.127707 -z)^2$ \\
        \textbf{5} & \textbf{0.30} & $\textbf{y = 28.42/ (1.68}^\textbf{z}\textbf{)}$ \\
        $7$ & $0.25$ & $y = \left(3.77 - z\right)^{2.32} + 6.24$ \\
        $8$ & $0.24 $ & $y = \left(\left(z - 3.18\right)^{2} - -3.47\right)\times2.06$ \\
        $9$ & $0.21$ & $y = {2.70}^{z} - 11.8\times z + 26.0$\\
        $11$ & $0.17$ & $y =4.19 \times\left(1 - 0.787\times z\right)^{6} + \left(4.94 - z\right)^{2.1}$ \\
        $13$ & $0.02$ & $y = \left(3.68 - z\right)^{2.4} + 6.56 - - \frac{0.02}{z- 0.76}$ \\
        $15$ & $0.016$ & $y = \left(3.68 - z\times\right)^{2.37} + 6.56 - - \frac{0.0169}{0.862 \left(z - 0.763\right)}$ \\
        $16$ & $0.015$ & $y = 21.9\times\left(1 - 0.272\times z\right)^{2.37} + 6.56 + \frac{0.0169}{\sqrt{z}\times \left(z - 0.763\right)}$ \\
        $17$ & $0.013$ & $y = \left(3.68 - z\right)^{2.37} + 6.56 - - \frac{0.0169}{z \times\left(z - 0.763 - 0.0216\right)}$ \\
        \hline
    \end{tabular}
    }
    \label{table:DESI-only}
\end{table}
\begin{table}[!ht]
    \centering
    \caption{DESI + Pantheon+.}
    \resizebox{\columnwidth}{!}{%
    \begin{tabular}{|c|c|l|}
        \hline
        \textbf{Complexity} & \textbf{Loss} & \textbf{Equation} \\ \hline
        $1$ & $124.5$ & $y = 1.24$ \\
        $2$ & $116.74$ & $y = e^{z}$ \\
        $3$ & $4.65$ & $y = 27.1\times z$ \\
        $5$ & $2.69$ & $y = \left(z/0.034\right)^{0.915}$ \\
        $6$ & $2.39$ & $y = z \times 30.20\times(0.18\times z -1)^2$ \\
        \textbf{7} & $\textbf{1.7}$ & $\textbf{y = (0.76}^\textbf{2} \times \textbf{z)/0.03}$ \\
        $9$ & $1.68$ & $y = 0.77^z\times(z\times29.96)^{0.99}$ \\
        $10$ & $1.68$ & $y = z^{3} + z \times\left(29.8 - z\times 8.24\right)$ \\
        $13$ & $1.68$ & $y =(z^{3} - z^{2} \times 7.60 + z\times 27.1)/{0.907}$ \\
        $14$ & $1.68$ & $y ={0.739}^{z} z \times\left(z \times\left(\sqrt{z} - 0.528\right) + 30.0\right)$ \\
        $17$ & $1.68$ & $y =z \times 30.0\times {1}/{\left(0.196 \times 0.465 + e^{0.266}\right)^{z}} + z^{2.13}$ \\
        $19$ & $1.68$ & $y =(30.0\times z)/{\sqrt{{1.95}^{z}}} + z^{2.13}$ \\\hline
    \end{tabular}
    }
    \label{table:DESI+PantheonPlus}
\end{table}
\begin{table}[!ht]
    \centering
    \caption{DESI + DESY5.}
    \resizebox{\columnwidth}{!}{%
    \begin{tabular}{|c|c|l|}
        \hline
        \textbf{Complexity} & \textbf{Loss} & \textbf{Equation} \\
        \hline
        $1$ & $796.$ & $y = z$ \\
        $2$ & $499.$ & $y = 8.65$ \\
        $3$ & $22.5$ & $y = 21.5\times z$ \\
        $5$ & $4.82$ & $y = \left(z \times 54.8\right)^{0.77}$ \\
        $6$ & $4.82$ & $y = 22.1\times z^{0.774}$ \\
        $\textbf{7}$ &\textbf{2.23} & $\textbf{y}\textbf{}= \textbf{(30.1}\times \textbf{z)/({1.31}}^\textbf{z}\textbf{)}$ \\
        $8$ & $2.10$ & $y = 29.44^z\times z $ \\
        $9$ & $2.05$ & $y = z\times \left(z+{30.1}/{{1.37}^{z}}\right)$ \\
        $11$ & $2.05$ & $y = z \times\left(z + {30.0}/{{1.37}^{z}}\right) - -0.0361$ \\
        $13$ & $2.05$ & $y = {z \times\left(z^{2} + 31.7\right)}/{\sqrt{{1.88}^{z}}} - z$ \\
        $14$ & $2.04$ & $y = 0.0129 +{z \times\left(\left(z^{2}\right)^{0.923} + 30.5\right)}/{{1.38}^{z}}$ \\
        $18$ & $2.04$ & $y ={31.1\times z \times\sqrt[4]{z^{0.2\times z}}}/{\sqrt{{1.85}^{z}}}$ \\
        $20$ & $2.04$ & $y = z\times \left(z + 30^{- z}\right) - {0.0218}/(z - 1.61)$ \\
        \hline
    \end{tabular}
    }
    \label{table:DESI+DESY5}
\end{table}
For $D_M / r_d$, the expressions from DESI+Pantheon and DESI+DESY5 appear different but are structurally equivalent. Both follow the form:
\begin{equation}
    \frac{D_M}{r_d} = a  z  \exp(bz).
\end{equation}
For DESI+Pantheon:
\begin{equation}
    \frac{D_M}{r_d} = \frac{0.76^z  z}{0.03} = 33.3  z  \exp(-0.274z),
\end{equation}
and for DESI+DESY5:
\begin{equation}
    \frac{D_M}{r_d} = \frac{30.1  z}{1.30^z} = 30.1  z  \exp(-0.262z).
\end{equation}
The small differences in coefficients reflect differences in the SNIa datasets.

Finally, PySR uses stochastic evolutionary methods, so small variations in the output can occur even when modeling the same quantity. This is expected due to the trade-off between accuracy and simplicity in symbolic regression \citep{cranmer2023interpretable}.
\section{Varying cosmological parameters}\label{append:gamma} 

In a complementary analysis of DESI + SNIa, we performed a test to explore the dependence of $w(z)$ on the parameters $\Omega_m$ and $r_d$ by varying their values in the calculation of $\gamma$. Specifically, we consider $\Omega_m = 0.25$, $0.30$, and $0.35$, as well as $r_d = 90.9 \,h^{-1} \, \text{Mpc}$, $101 \,h^{-1} \, \text{Mpc}$, and $111.1 \,h^{-1} \, \text{Mpc}$ -- chosen to lie below and above the fiducial values adopted in this work, although still compatible with most of their measurements in the literature.

\begin{figure}[!ht]
    \includegraphics[scale=0.3]{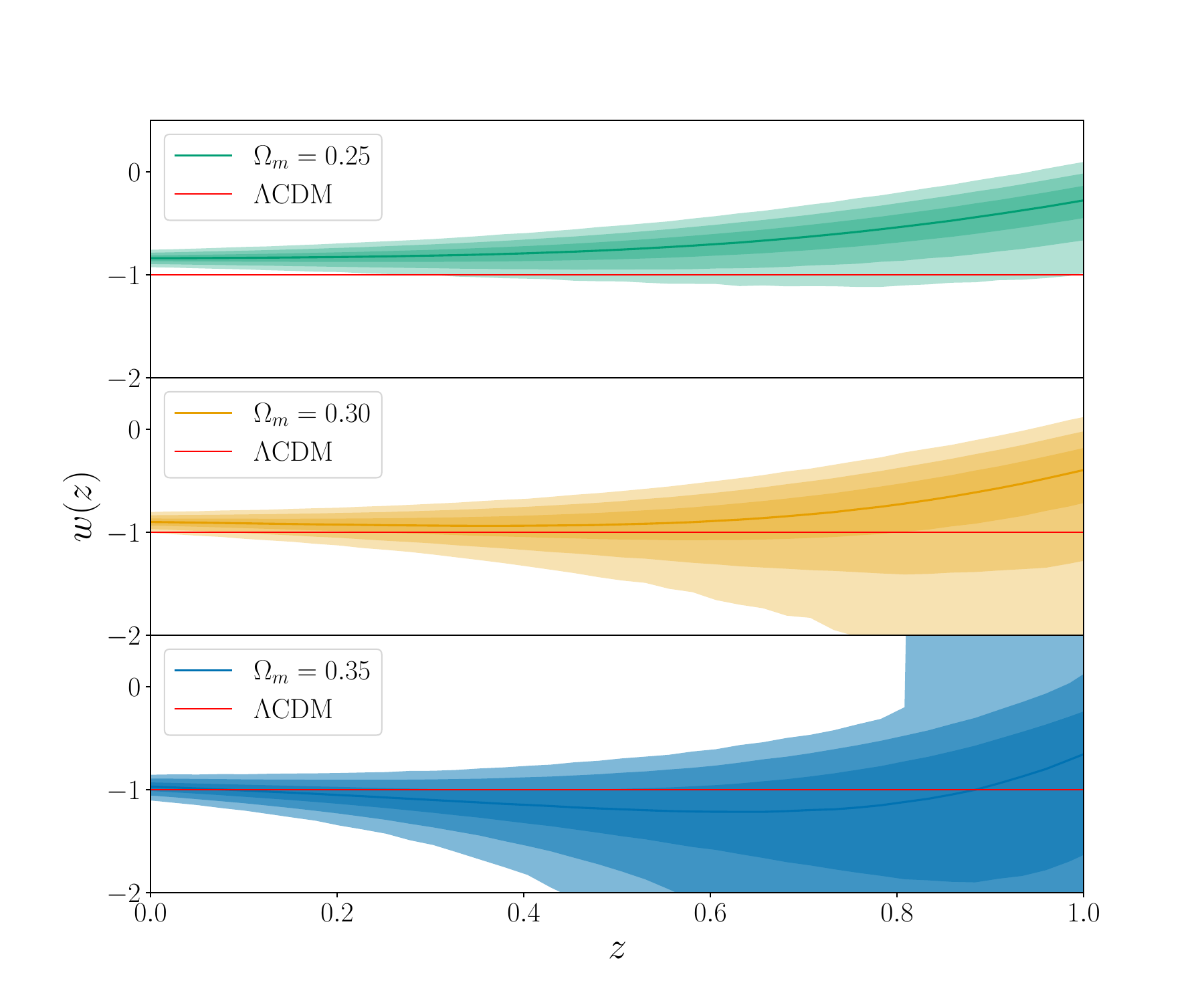}
    \includegraphics[scale=0.3]{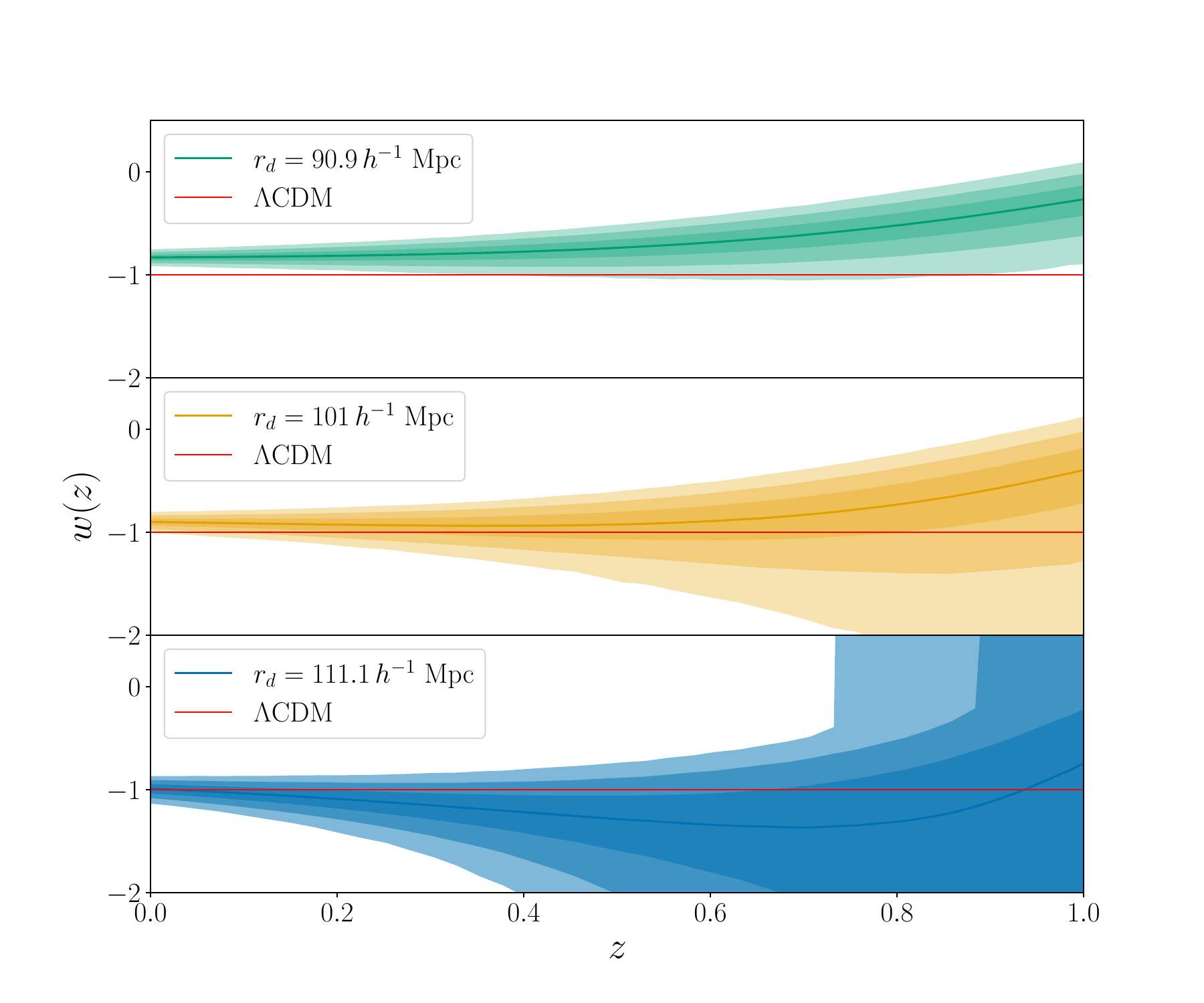}

\caption{Reconstruction ($3\sigma$) of $w(z)$ for DESI + Pantheon+ considering variations in $\Omega_m$ and $r_d$ for different scenarios. 
        \textbf{Top:} $w(z)$ reconstruction with fixed $r_d$. 
        \textbf{Bottom:} $w(z)$ reconstruction with fixed $\Omega_m$, for the same scenario.}
    \label{fig:panth_variat}
\end{figure}

\begin{figure}[!ht]
    \centering
    \includegraphics[scale=0.3]{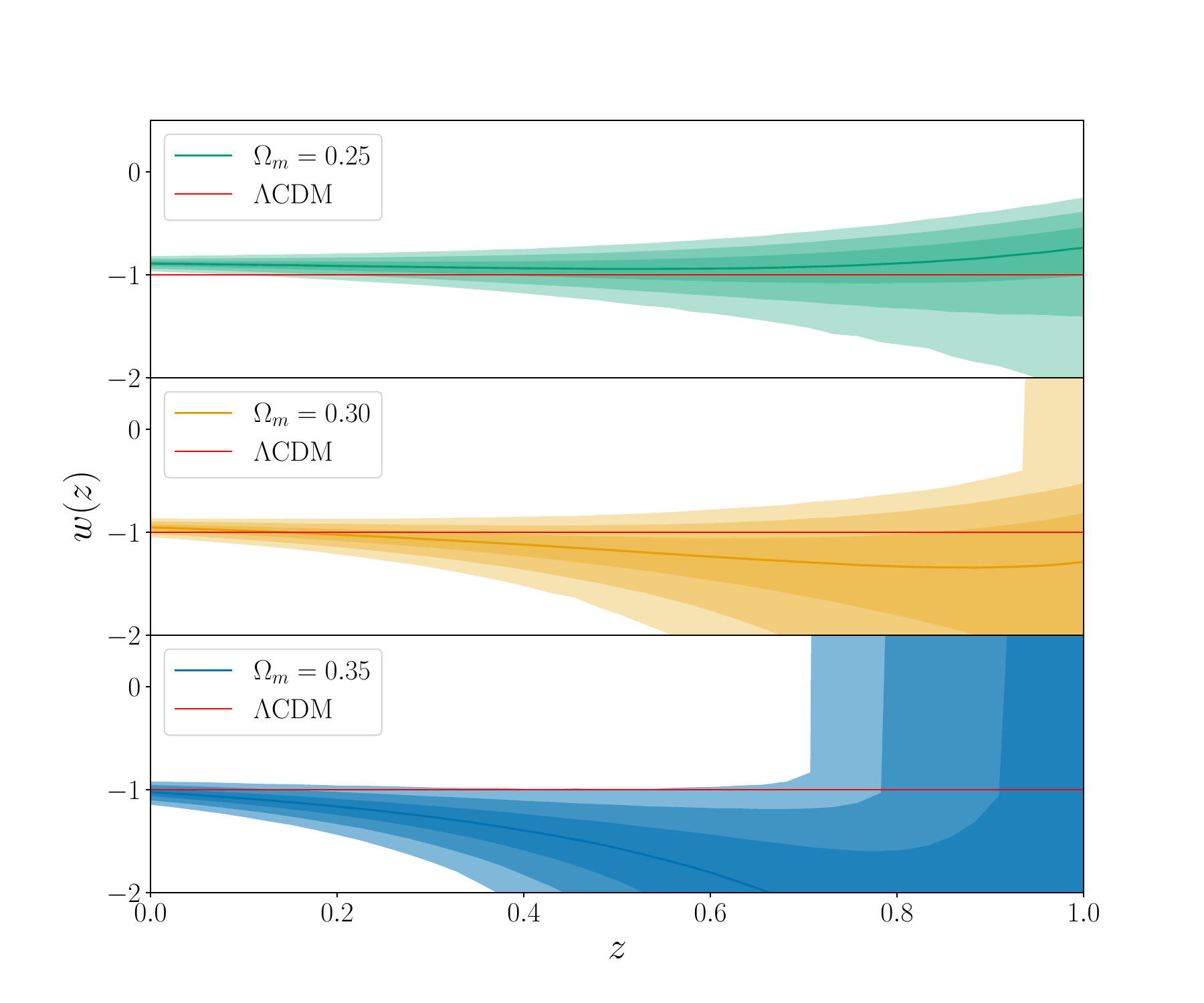}
    \includegraphics[scale=0.3]{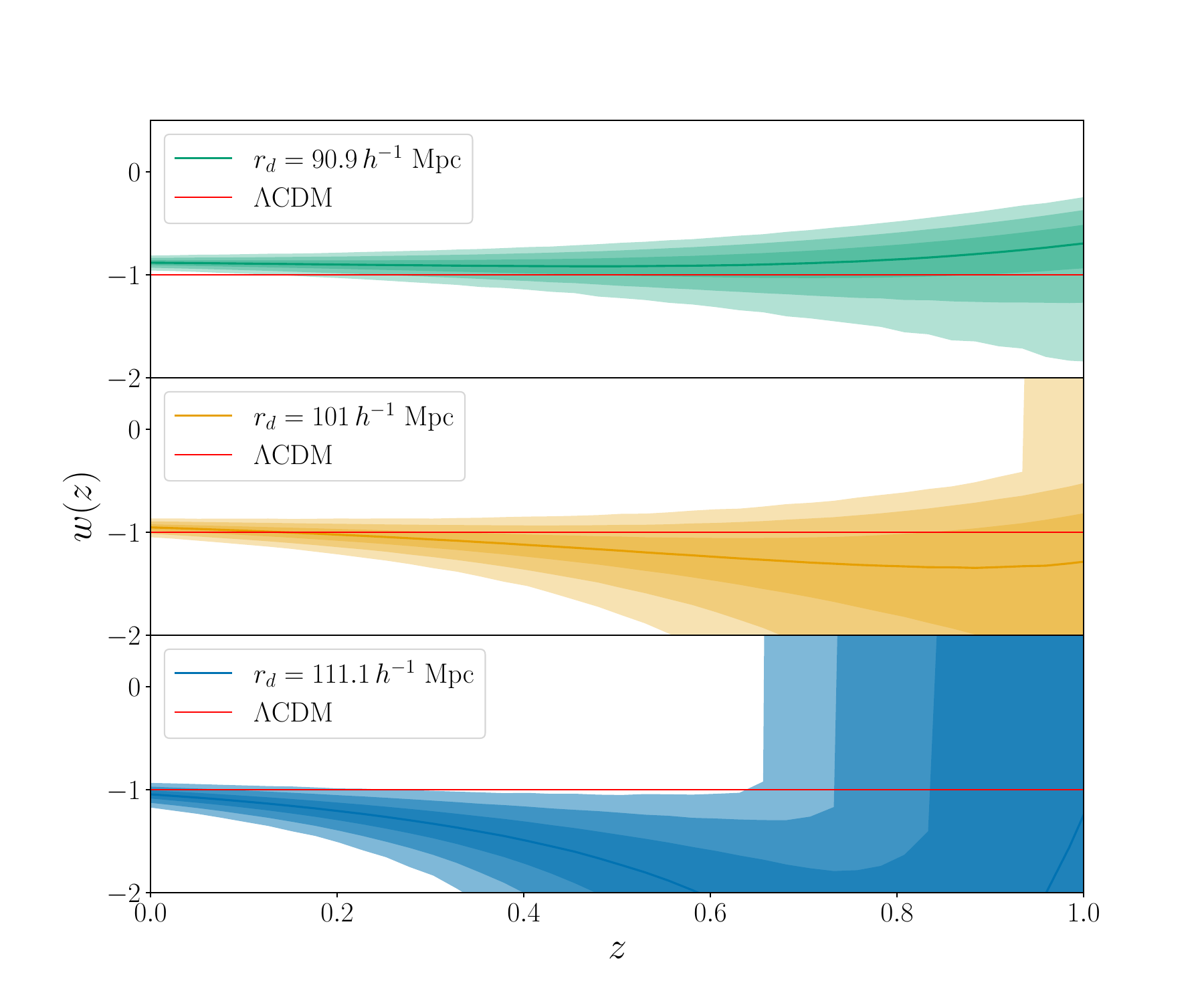}
\caption{The same as the Figure \ref{fig:panth_variat} for DESI + DESY5.}
    \label{fig:DESY5_variat}
\end{figure}

The results are presented in Figures \ref{fig:panth_variat} and ~\ref{fig:DESY5_variat}. In the upper panel of Figure \ref{fig:panth_variat}, we analyze the scenario of DESI + Pantheon, fixing $r_d = 101\,h^{-1} \, \text{Mpc}$ while varying different values of $\Omega_m$. In the bottom panel of Figure \ref{fig:panth_variat}, we fix $\Omega_m = 0.30$ and allow $r_d$ to vary. Figure~\ref{fig:DESY5_variat} also contains two panels: the upper panel shows the reconstruction of $w(z)$ with $r_d$ fixed for the DESI + DESY5 scenario, allowing $\Omega_m$ to vary. The bottom panel presents $w(z)$ with $\Omega_m$ fixed while varying $r_d$.

In all panels, the colored regions indicate the 3$\sigma$ confidence intervals obtained from MCM, and the black dashed line represents the theoretical $\Lambda$CDM curve ($w = -1$). These results demonstrate the sensitivity of $w(z)$ to the cosmological parameters in Eq. (\ref{dinda&Maartins}). It is notable that mild deviations from the 3$\sigma$ (C.L.) occur when the smallest values of $\Omega_m$ and $r_d$ are assumed. These lower values represent more extreme choices compared to the higher values, which are better supported by current data, indicating that our results remain robust under reasonable variations in parameters.

\section{Analyses with DESI DR1}

In what follows, we present the reconstructed results obtained using the method described in Section~\ref{CF}, based on DESI DR1 data. These results are shown here for completeness and comparison, as the main analysis in this work has been updated to use DESI DR2. 

First, we combine DESI DR1 with Pantheon+, and subsequently with DESY5. The {best} equation derived by PySR for the $D_M/r_d$ reconstruction, in the first scenario, is given by 
\begin{equation}\label{bestPantheonDR1}
  \frac{D_M}{r_d} = 29.8 \times0.760^{z} z\;.
\end{equation}
Overall, the reconstruction using DR1 is consistent with that obtained from DR2. No significant changes are observed in the reconstructed trends, with both versions showing good agreement with the $\Lambda$CDM model.

Related to the $w(z)$ reconstruction for DESI DR1 + Pantheon+, no significant deviations are observed when compared to the DR2-based reconstruction, indicating that the transition from DR1 to DR2 did not introduce notable changes in the inferred behavior of $ w(z)$.
\begin{figure}[!ht]
    \centering
    \includegraphics[scale=0.3]{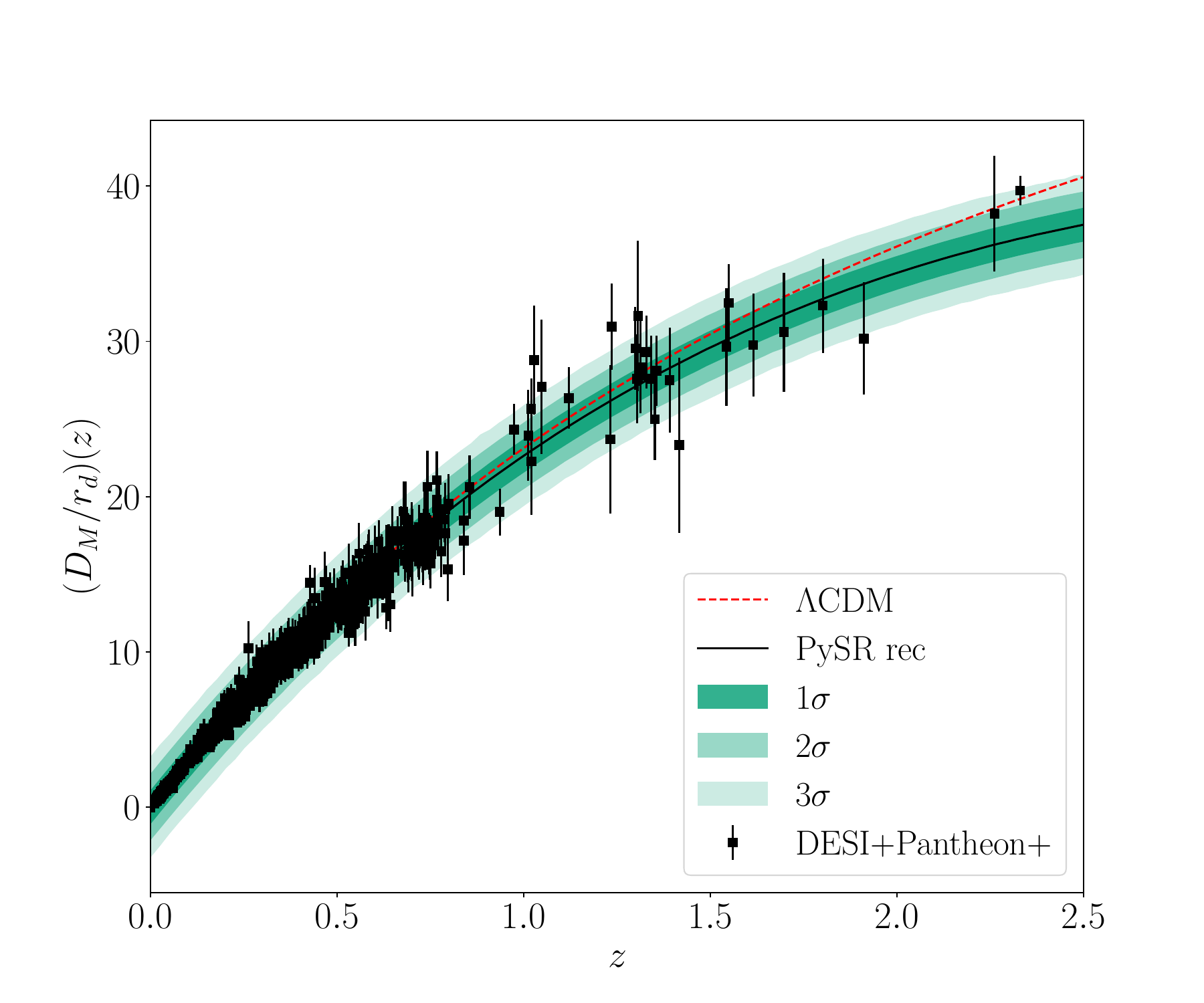}
    \includegraphics[scale=0.3]{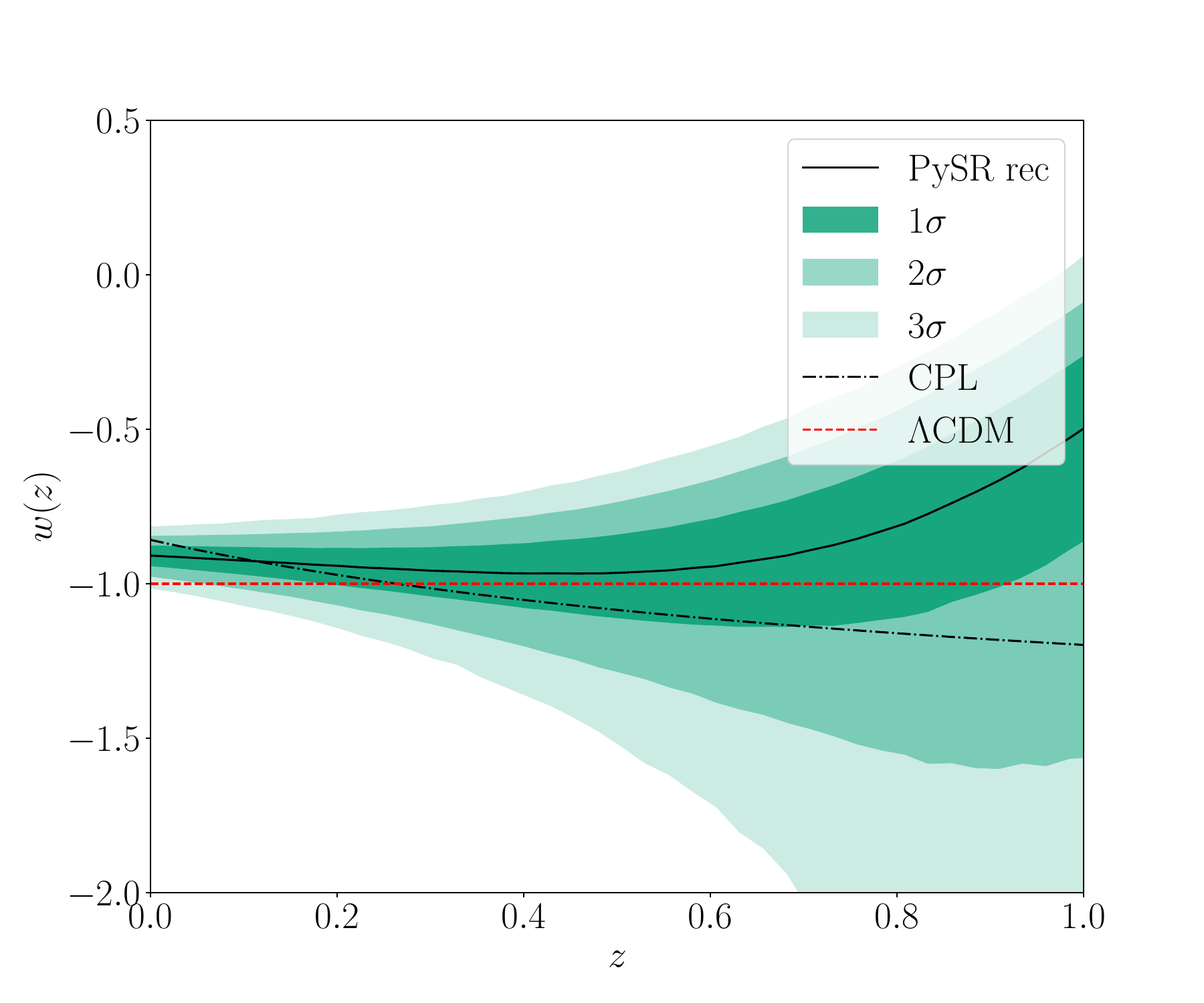}
\caption{The reconstruction results for $D_M/r_d$ using DESI + Pantheon+ dataset. The black line represents the mean for the PySR reconstruction; the green bands indicate the confidence levels. The red dashed line corresponds to the $\Lambda$CDM theoretical curve. \textbf{Top:} $D_M/r_d$ reconstruction with PySR. The black points correspond to the dataset, with respective uncertainties. 
\textbf{Bottom:} The $w(z)$ derived using Eq.~\eqref{dinda&Maartins2} and the reconstructed $D_M/r_d$, assuming $r_d=101\pm2.3h^{-1}$ Mpc, as reported by \cite{Verde_2017}, and $\Omega_m=0.3$. While the black dashed line represents the CPL model with the DESI (FS + BAO) + CMB + Pantheon+ results \cite{desicollaboration2024desi2024viicosmological}.}
\label{fig:pantheonre}
\end{figure}

\begin{figure}[!ht]
    \centering
    \includegraphics[scale=0.3]{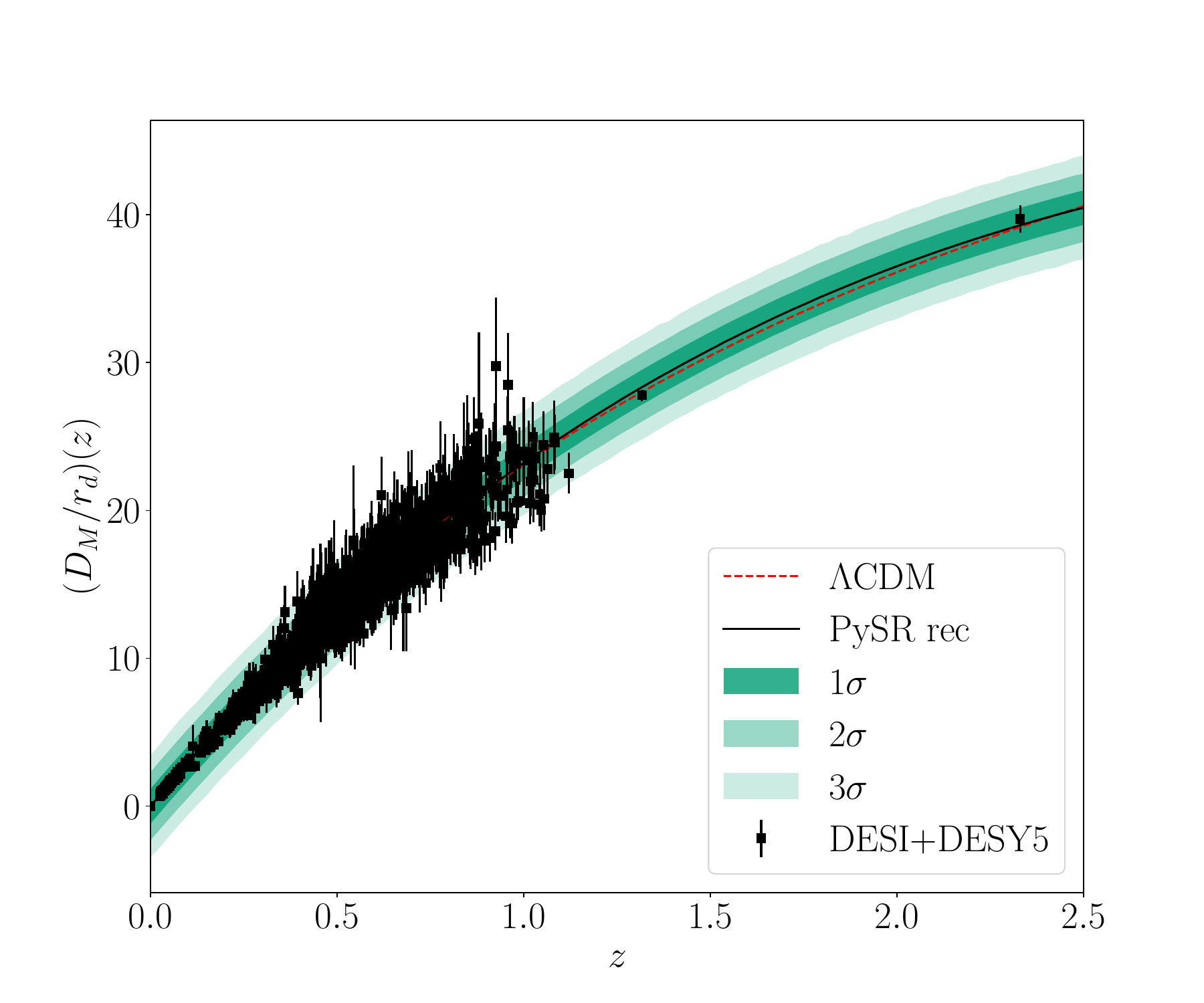}
    \includegraphics[scale=0.3]{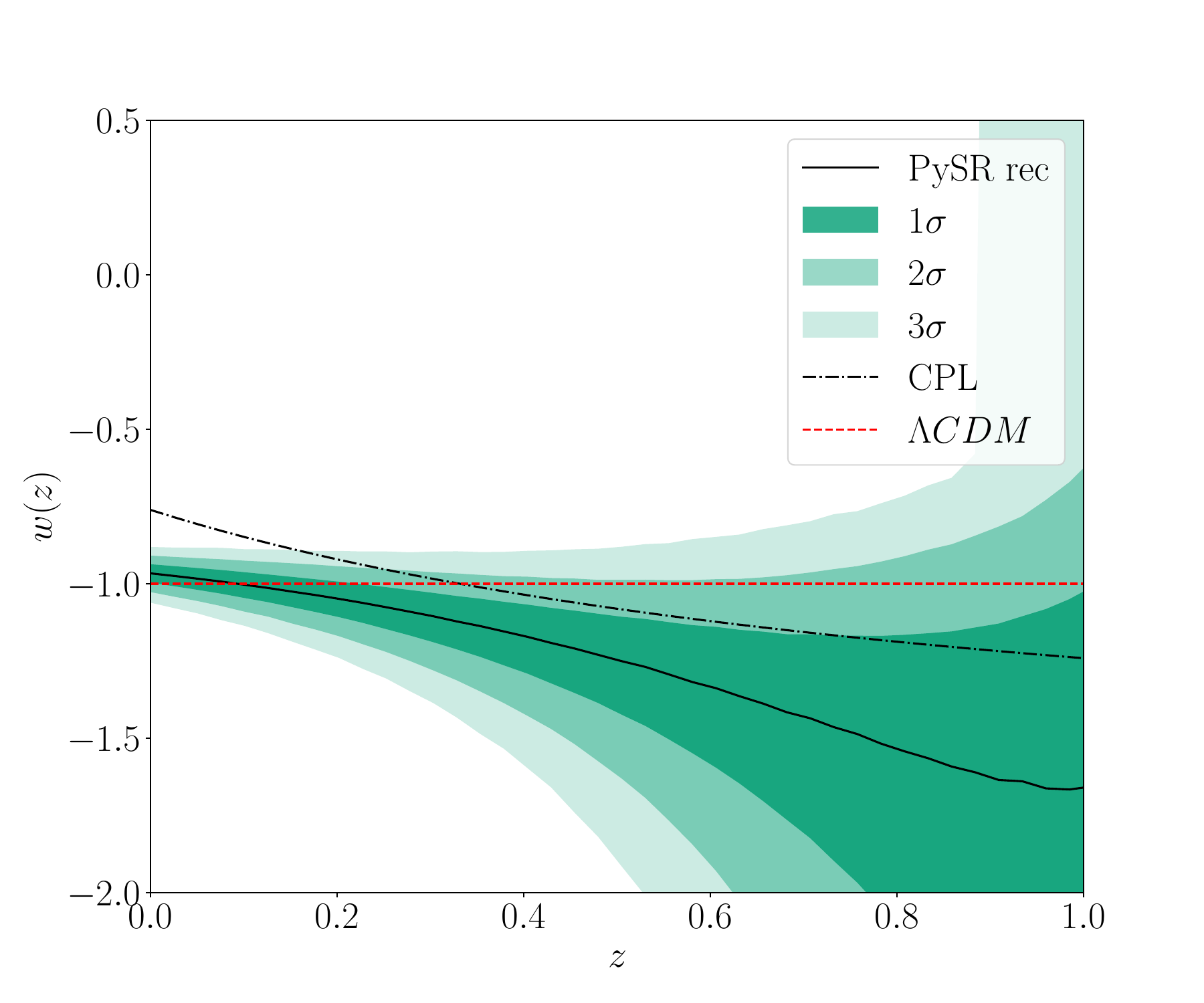}
\caption{ The same as Figure~\ref{fig:pantheonre}, but using the DESI + DESY5, and a CPL model with the DESI (FS + BAO) + CMB + DESY5 results \cite{desicollaboration2024desi2024viicosmological}.}
    \label{fig:DESY5H070}
\end{figure}

The resulting data for DESI DR1 + DESY5 is used to derive a new relationship for $D_M / r_d$. The best equation obtained from PySR for this dataset is
\begin{equation}\label{bestDESY5}  
\frac{D_M}{r_d} = 29.8\times{0.784}^{z}z\;. 
\end{equation}  
In this case, we observe good agreement with the $\Lambda$CDM model across most of the redshift range for the $D_M / r_d$ reconstruction. Notably, for $z > 1.5$, the reconstruction with DESI DR1 + DESY5 remains closer to $\Lambda$CDM compared to the DESI DR2 results, which show a mild deviation in that region (see Figure~\ref{fig:DESY5H070}, top). The reconstructed $w(z)$ shows a distinct evolution compared to the result from DESI DR2. At high redshifts, $w(z) < -1$, increasing with time and crossing the phantom behaviours around $z \sim 0.2$, reaching $w(z) > -1$ today ($z = 0$) (see Figure~\ref{fig:DESY5H070}, bottom).

\FloatBarrier
\bibliography{apssampV_2}

\end{document}